\documentstyle[12pt]{article}
\newlength{\extraspace}
\newlength{\extraspaces}
\newcommand{\be}{\begin{equation}
\addtolength{\abovedisplayskip}{\extraspaces}
\addtolength{\belowdisplayskip}{\extraspaces}
\addtolength{\abovedisplayshortskip}{\extraspace}
\addtolength{\belowdisplayshortskip}{\extraspace}}
\newcommand{\ee}{\end{equation}}

\newcommand{\ba}{\begin{eqnarray}
\addtolength{\abovedisplayskip}{\extraspaces}
\addtolength{\belowdisplayskip}{\extraspaces}
\addtolength{\abovedisplayshortskip}{\extraspace}
\addtolength{\belowdisplayshortskip}{\extraspace}}
\newcommand{\ea}{\end{eqnarray}}

\newcommand{\newsection}[1]{
\vspace{15mm}
\pagebreak[3]
\addtocounter{section}{1}
\setcounter{equation}{0}
\setcounter{subsection}{0}
\setcounter{footnote}{0}
\begin{flushleft}
{\large\bf \thesection. #1}
\end{flushleft}
\nopagebreak
\medskip
\nopagebreak}

\newcommand{\Tr}{{\rm Tr}}
\newcommand{\Dmrns}{{\cal D}_{\mu\rho,\nu\sigma}}

\begin{document}

\addtolength{\baselineskip}{.8mm}

{\thispagestyle{empty}
\noindent \hspace{1cm}  \hfill IFUP--TH/2002--10 \hspace{1cm}\\
\mbox{}                 \hfill Revised version \hspace{1cm}\\

\begin{center}\vspace*{1.0cm}
{\large\bf On the dependence of the gauge--invariant field--strength} \\
{\large\bf correlators in QCD on the shape of the Schwinger string
 \footnote{Partially supported by MIUR (Italian Ministry of the University 
 and of Scientific and Technological Research) and by the INTAS contract
 00--0110.} }\\
\vspace*{1.0cm}
{\large A. Di Giacomo, E. Meggiolaro}\\
\vspace*{0.5cm}{\normalsize
{Dipartimento di Fisica, \\
Universit\`a di Pisa, \\ 
and INFN, Sezione di Pisa,\\
I--56100 Pisa, Italy.}}\\
\vspace*{2cm}{\large \bf Abstract}
\end{center}
\noindent
We study, by numerical simulations on a lattice, the dependence of the
gauge--invariant two--point field--strength correlators in QCD on the
path used to perform the color parallel transport between the two points.\\
\vspace{1.0cm}
\noindent
(PACS code: 12.38.Gc)
}
\vfill\eject

\newsection{Introduction}

\noindent
The gauge--invariant two--point correlators of the field strengths in the 
QCD vacuum are defined as
\be
\Dmrns^{({\cal C})}(x) = g^2 \langle 0| 
\Tr \left\{ G_{\mu\rho}(0) S(0,x | {\cal C}) G_{\nu\sigma}(x)
S^\dagger(0,x | {\cal C}) \right\} |0\rangle ~,
\label{corr}
\ee
where $G_{\mu\rho} = T^a G^a_{\mu\rho}$ and $T^a$ are the matrices of the 
algebra of the color group SU(3) in the fundamental representation 
The trace in (\ref{corr}) is taken on the color indices.
\be
S(0,x | {\cal C}) \equiv {\rm P} \exp \left( i g \int_{0,~{\cal C}}^x
dz^\mu A_\mu (z) \right) ~,
\label{string}
\ee
with $A_\mu = T^a A^a_\mu$, is the Schwinger phase operator needed to 
parallel--transport the tensor $G_{\nu\sigma}(x)$ to the point $0$.
``P'' stands for ``path ordering'' and ${\cal C}$ is any path from $0$
to $x$.

Field--strength correlators play an important role in hadron physics.
In the spectrum of heavy $Q \bar{Q}$ bound states, they govern the effect 
of the gluon condensate on the level splittings 
\cite{Gromes82,Campostrini86,Simonov95}.
They are the basic quantities in models of stochastic confinement of color
\cite{Dosch87,Dosch88,Simonov89}
and in the description of high--energy hadron scattering \cite{Nachtmann84,
Landshoff87,Kramer90,Dosch94,Nachtmann97,Nachtmann99,Pirner02}.
In some recent works \cite{Dorokhov97,Ilgenfritz98}, these correlators have 
been semi--classically evaluated in the single--instanton approximation and
in the instanton dilute--gas model, so providing useful information about the 
role of the semiclassical modes in the QCD vacuum.

The correlators (\ref{corr}) in principle depend on the choice of the path
${\cal C}$. Usually, in developing the ``Stochastic Vacuum Model'' (SVM), the
path ${\cal C}$ appearing in Eqs. (\ref{corr}) and (\ref{string}) is assumed
to be the straight line connecting the points $0$ and $x$. Lattice data
\cite{Campostrini84,DiGiacomo92,npb97,plb97},
which are the input for these SVM calculations, also refer to the choice of
the straight--line parallel transport. With this choice the Schwinger string
(\ref{string}) reads:
\be
S(0,x) = {\rm P}\exp\left(i g \int^1_0dt\,x^\mu A_\mu(xt)\right) ~.
\label{line}
\ee
Then, in the Euclidean region, translational, O(4)-- and parity invariance
require the correlator (\ref{corr}) to be of the following form 
\cite{Dosch87,Dosch88,Simonov89}:
\ba
\lefteqn{
\Dmrns(x) = (\delta_{\mu\nu}\delta_{\rho\sigma} - 
\delta_{\mu\sigma}\delta_{\rho\nu})
\left[ {\cal D}(x^2) + {\cal D}_1(x^2) \right] } \nonumber \\
& & + (x_\mu x_\nu \delta_{\rho\sigma} - x_\mu x_\sigma \delta_{\rho\nu} 
+ x_\rho x_\sigma \delta_{\mu\nu} - x_\rho x_\nu \delta_{\mu\sigma})
{\partial{\cal D}_1(x^2) \over \partial x^2} ~,
\label{param}
\ea
where ${\cal D}$ and ${\cal D}_1$ are invariant functions of $x^2$.

The functions ${\cal D}(x^2)$ and ${\cal D}_1(x^2)$ have been directly 
determined by numerical simulations on a lattice in
the {\it quenched} (i.e., pure--gauge) theory, with gauge group SU(2) 
\cite{Campostrini84}, in the {\it quenched} SU(3) theory in the range of 
physical distances between 0.1 and 1 fm \cite{DiGiacomo92,npb97} and also in 
full QCD, i.e., including the effects of dynamical fermions \cite{plb97}.

No analysis exists of what happens for a different choice of the path
${\cal C}$ than the straight line, except for qualitative statements that
the physics should not strongly depend on the deformations of the
path \cite{Dosch87,Dosch88,Simonov89}.
In this paper we compute the gauge--invariant field--strength correlators for
deformed paths (see Fig. 1), in the {\it quenched} SU(3) lattice gauge theory,
for the benefit of the developments of the SVM.

\newsection{Computations and results}

\noindent
In principle, every choice of a path ${\cal C}_{[0,x]}$ connecting the two
points $0$ and $x$ in the expression (\ref{string}) for the Schwinger string
operator will generate a different field--strength correlator,
that we have denoted in Eq. (\ref{corr}) as $\Dmrns^{({\cal C})}(x)$.

On the lattice we can define a lattice operator $\Dmrns^{({\cal C})L}$, which
is proportional to $\Dmrns^{({\cal C})}$ in the continuum limit, when the
lattice spacing $a \to 0$. Since the lattice analogue of the field strength is
the open plaquette $\Pi_{\mu\rho}(n)$ (the parallel transport along an
elementary square of the lattice lying on the $\mu\rho$--plane, starting from
the lattice site $n$ and coming back to $n$), $\Dmrns^{({\cal C})L}$ will be
defined as \cite{DiGiacomo92,npb97,plb97}
\be
\Dmrns^{({\cal C})L}(\hat d a) = {1\over2} \Re \left\{ \langle \Tr [
\Pi_{\mu\rho}(n) S(n,n + {\hat d}a | {\cal C}) 
\Omega_{\nu\sigma}(n + {\hat d}a)
S^\dagger(n,n + {\hat d}a | {\cal C}) \rangle \right\} ~,
\label{corr-latt}
\ee
where $\Re$ stands for real part and 
the lattice operator $\Omega_{\nu\sigma}(n)$ is given by
\cite{DiGiacomo92,npb97}
\be
\Omega_{\nu\sigma}(n) =
\Pi^\dagger_{\nu\sigma}(n) - \Pi_{\nu\sigma}(n)
-{1 \over 3} \Tr [\Pi^\dagger_{\nu\sigma}(n) - \Pi_{\nu\sigma}(n)] ~.
\label{omega}
\ee
As explained in Refs. \cite{DiGiacomo92,npb97}, the inclusion of the operator
${1 \over 2} \Omega_{\nu\sigma}$ (in place of $\Pi^\dagger_{\nu\sigma}$)
on the right--hand side of Eq. (\ref{omega})
ensures that the {\it disconnected} part and the {\it singlet} part of the
correlator are left out, so that we are indeed taking the correlation of two 
operators with the exchange of the quantum numbers of a color octet.

The lattice site $n + {\hat d}a$ is the site at a distance $d$ lattice
spacings from $n$, in the direction of a coordinate axis.
The Schwinger string $S(n,n+{\hat d}a | {\cal C})$ is a parallel
transport connecting the point $n$ with the point $n + {\hat d}a$ along the
path ${\cal C}$.

We have explored the dependence on the path by measuring the correlators
for the various paths shown in Fig. 1.
Fig. 1a shows the straight--line path [corresponding to the straight--line
Schwinger operator in Eq. (\ref{line})], which, as explained in the
introduction, has already been widely analyzed.
Fig. 1b shows our second choice: the {\it transverse} size $\delta$, in units
of the lattice spacing $a$, parametrizes the deviation from the straight--line
case. We shall label by $b_\delta$ the paths in
Fig. 1b, for a {\it given} value of the transverse size $\delta$.
If an average is made over the orientation of the staple in the plane
orthogonal to ${\hat d}$, the correlator will have the same symmetries
(translation, O(4) and parity)
as the one in Fig. 1a and the parametrization (\ref{param}) will still
apply, with, of course, different functions ${\cal D}^{(b_\delta)}$
and ${\cal D}^{(b_\delta)}_1$.
The same considerations also apply to the {\it collection} $c_\delta$ of paths
shown in Fig. 1c, for a given value of the transverse size $\delta$,
if we average over the orientation of the plane containing the path.
We have chosen the {\it inversion point} $p$ in Fig. 1c just in the
middle of the line $(0,x)$, i.e., p = x/2, in order to preserve parity
invariance. For other choices, i.e., $p = \alpha x$ with $\alpha \in [0,1]$
and $\alpha \not= 1/2$, one should average between the path with
$p = \alpha x$ and the path with $p = (1-\alpha)x$ in order to preserve parity.

In both cases, we define \cite{DiGiacomo92} the
two independent functions ${\cal D}^{({\cal C})}_\parallel(x^2)$ and
${\cal D}^{({\cal C})}_\perp(x^2)$ as follows.
We go to a reference frame in which $x^\mu$ is parallel to one of the
coordinate axes, say $\mu=0$. Then
\ba
{\cal D}^{({\cal C})}_\parallel &\equiv& {1\over3} \sum_{i=1}^3
{\cal D}^{({\cal C})}_{0i,0i}(x)
= {\cal D}^{({\cal C})} + {\cal D}^{({\cal C})}_1
+ x^2 {\partial{\cal D}^{({\cal C})}_1 \over \partial x^2} ~, \nonumber \\
{\cal D}^{({\cal C})}_\perp &\equiv& {1\over3}\sum_{i<j=1}^3
{\cal D}^{({\cal C})}_{ij,ij}(x)
= {\cal D}^{({\cal C})} + {\cal D}^{({\cal C})}_1 ~,
\label{para-perp}
\ea
where ${\cal D}^{({\cal C})} = {\cal D}^{({\cal C})}(x^2)$ and
${\cal D}^{({\cal C})}_1 = {\cal D}^{({\cal C})}_1 (x^2)$ are the two
invariant functions entering in the parametrization (\ref{param}) for
the given {\it collection} of paths ${\cal C}$.

In the na\"\i ve continuum limit ($a \to 0$)
\be
\Dmrns^{({\cal C})L}({\hat d}a) \mathop\sim_{a\to0} a^4 \Dmrns^{({\cal C})}
({\hat d}a) + {\cal O}(a^6) ~.
\label{corr-limit}
\ee
Making use of the definition (\ref{para-perp}) we can also write,
in the same limit,
\ba
{\cal D}^{({\cal C})L}_\parallel(d^2 a^2) \mathop\sim_{a\to0} a^4 
{\cal D}^{({\cal C})}_\parallel(d^2 a^2) + {\cal O}(a^6) ~,\nonumber \\
{\cal D}^{({\cal C})L}_\perp(d^2 a^2) \mathop\sim_{a\to0} a^4 
{\cal D}^{({\cal C})}_\perp(d^2 a^2) + {\cal O}(a^6) ~.
\label{para-perp-limit}
\ea
In order to remove the {\it lattice artefacts}, i.e., the terms
${\cal O}(a^6)$ in Eqs. (\ref{corr-limit}) and (\ref{para-perp-limit}),
we shall make use of the {\it cooling} technique, described in previous
papers (see Refs. \cite{Campostrini89,DiGiacomo90}
and \cite{DiGiacomo92,npb97,plb97}).
Cooling is a local procedure, which affects correlations at distances that
grow as the square root of the number of cooling steps, as in 
a diffusion process.
If the distance at which we observe the correlation is sufficiently large,
we then expect that lattice artefacts are frozen by cooling long before the
correlation is affected: this will produce a {\it plateau} in the dependence
on the cooling step.
Our data are the values of the correlators at the {\it plateau}; 
the error is the typical statistical error at the {\it plateau}, 
{\it plus} a systematic error which is estimated as the difference 
between neighbouring points at the {\it plateau}.\footnote[2]{In Figs. 2, 3, 4
and 5 we have plotted only the points corresponding to a clear {\it plateau}
in the cooling process. For some points at large distances, our cooling
proved to be not long enough to reach the {\it plateau}.}

We have measured the correlations on a $16^4$ lattice at distances ranging 
from 3 to 8 lattice spacings and at $\beta = 6$. The value of the lattice
spacing $a$ in physical units can be extracted from the value of the string
tension \cite{Michael88,Bali93} and it is $a \simeq 0.1$ fm at our value
of $\beta$. At this value of $\beta$ scaling has been already tested for
the case of the straight--line paths in Refs. \cite{DiGiacomo92,npb97}.
The data of Refs. \cite{DiGiacomo92,npb97} came from several different
values of $\beta$ (including also $\beta = 6$) and two lattice sizes,
$16^4$ and $32^4$. They showed no visible dependence neither on the
ultraviolet, nor on the infrared scale.
We therefore do not expect significant scaling violations.

The results are shown in Figs. 2, 3, 4 and 5, versus the distance $d$ in units
of the lattice spacing $a$. The lines are drawn as an eye--guide.
Figs. 2 and 3 refer to the paths in Fig. 1b,
for various values of the {\it transverse} size $\delta$ of the staple.
Figs. 4 and 5 give the analogous results for the paths shown in Fig. 1c.

The very result of this paper is the unexpectedly strong dependence of the
correlators on the shape of the path.
The correlator with the straight--line path (\ref{line}) (see Fig. 1a),
has the largest signal (for every distance $d$), compared with the two other
choices for the path. Every deformation from the straight--line path
seems to produce a sharp decrease of the value of the correlator.
What seems to be path--independent is the slope of the curves, which
look parallel to one another in the linear--log plots of Figs. 2, 3,
4 and 5. This means that the correlation length $\lambda_A$ defined
in Refs. \cite{DiGiacomo92,npb97,plb97} is approximately path--independent,
at least for the classes of paths that we have considered.

In Figs. 2, 3, 4 and 5 the best fit to the data for straight--line Schwinger
string of Ref. \cite{npb97}, Eqs. (2.10) and (2.11), is displayed for
comparison (see also Ref. \cite{EM99}, where a critical comparison among
different best fits is performed).

In order to quantitatively test the independence of the correlation length
$\lambda_A$ on the path ${\cal C}$, we have tried a best fit
to the data for ${\cal D}_\perp = {\cal D} + {\cal D}_1$, at intermediate
distances $5 \le d \le 8$, i.e., about $0.5~{\rm fm} \le x \le 0.8$ fm),
with the same function used in Ref. \cite{npb97}, Eq. (2.10):
\be
{\cal D}_\perp (x^2) = A_\perp {\rm e}^{-|x|/\lambda_A}
+ {a_\perp \over |x|^4} {\rm e}^{-|x|/\lambda_a} ~.
\label{fit-perp}
\ee
Keeping $\lambda_A$ fixed to the value of Ref. \cite{npb97}, Eq. (2.11),
i.e., $\lambda_A = 1 / 182 \Lambda_L$, where $\Lambda_L \simeq 4.9$ MeV
is the lattice scale \cite{Michael88,Bali93},
a good fit results with a reasonable $\chi^2/N_{d.o.f.}$.
The path--dependence of the correlator reflects in a rather strong
path--dependence of the coefficient $A_\perp$ of the exponential term.
In the range of distances chosen, the second term of Eq. (\ref{fit-perp})
is compatible with zero, within the errors.
The coefficient $A_\perp$ changes up to a factor of four, going from
$\delta=0$ to $\delta=6$ for the collection of paths $b_\delta$ (Fig. 1b);
and it changes up to a factor of two, going from $\delta=0$ to $\delta=6$
for the collection of paths $c_\delta$ (Fig. 1c).

In Ref. \cite{npb97} Eq. (\ref{fit-perp}) was considered as a split--point
regulator of the gluon condensate, which was extracted from the coefficient
$A_\perp$ of the exponential term (see also Ref. \cite{EM99}).
It is not clear to us how this determination depends on the shape of the path.

\bigskip
\noindent {\bf Acknowledgements}
\smallskip

This work was done using the CRAY T3E of the CINECA Inter University 
Computing Centre (Bologna, Italy). We would like to thank the CINECA for 
the kind and highly qualified technical assistance.
 
Many discussions on this subject with H.G. Dosch, H.J. Pirner and
Yu.A. Simonov are warmly acknowledged.

\vfill\eject

{\renewcommand{\Large}{\normalsize}
}

\vfill\eject

\noindent
\begin{center}
{\bf FIGURE CAPTIONS}
\end{center}
\vskip 0.5 cm
\begin{itemize}
\item [\bf Fig.~1.] The different paths for the Schwinger string $S$
that we have considered for our analysis.
\bigskip
\item [\bf Fig.~2.] The function $a^4 {\cal D}^{(b_\delta)}_\parallel$ versus
the distance $d$ in units of the lattice spacing $a$.
{\it Circles} correspond to the straight--line path of Fig. 1a; 
{\it triangles} to Fig. 1b with $\delta = 1$;
{\it squares} to Fig. 1b with $\delta = 4$;
{\it diamonds} to Fig. 1b with $\delta = 6$.
Lines are drawn as eye--guides.
The thick continuum line has been obtained using the parameters of the best fit
obtained in Ref. \cite{npb97}, Eqs. (2.10) and (2.11).
\bigskip
\item [\bf Fig.~3.] The function $a^4 {\cal D}^{(b_\delta)}_\perp$ versus
the distance $d$ in units of the lattice spacing $a$.
The symbols are the same as in Fig. 2.
\bigskip
\item [\bf Fig.~4.] The function $a^4 {\cal D}^{(c_\delta)}_\parallel$ versus
the distance $d$ in units of the lattice spacing $a$.
The symbols are the same as in Fig. 2, except that they now refer to 
the path of Fig. 1c.
\bigskip
\item [\bf Fig.~5.] The function $a^4 {\cal D}^{(c_\delta)}_\perp$ versus
the distance $d$ in units of the lattice spacing $a$.
The symbols are the same as in Fig. 2, except that they now refer to 
the path of Fig. 1c.
\end{itemize}

\vfill\eject

\pagestyle{empty}

\centerline{\bf Figure 1}
\vskip 4truecm
\begin{figure}[htb]
\vskip 4.5truecm
\includegraphics{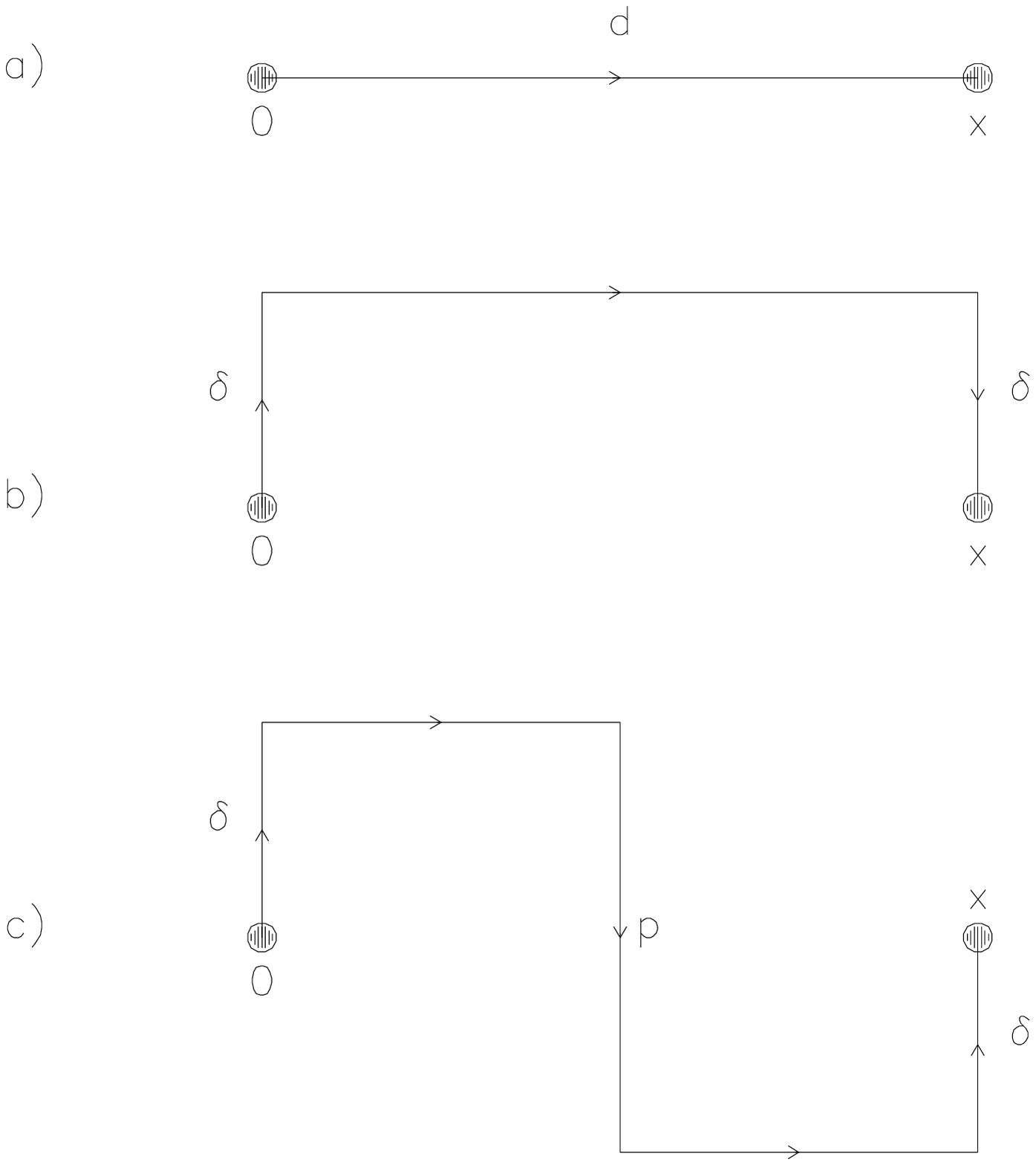}
\end{figure}

\vfill\eject

\centerline{\bf Figure 2}
\vskip 4truecm
\begin{figure}[htb]
\vskip 4.5truecm
\includegraphics{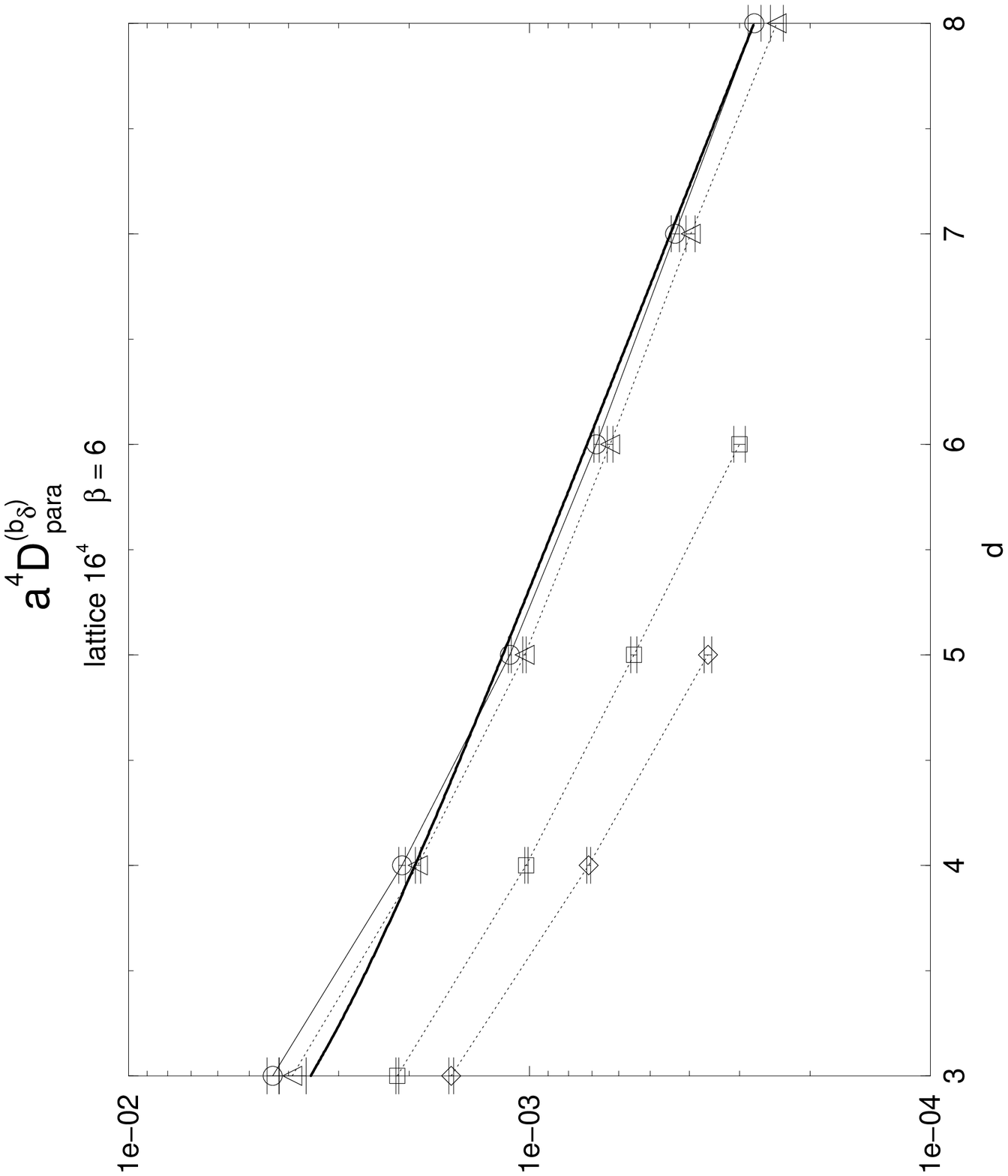}
\end{figure}

\vfill\eject

\centerline{\bf Figure 3}
\vskip 4truecm
\begin{figure}[htb]
\vskip 4.5truecm
\includegraphics{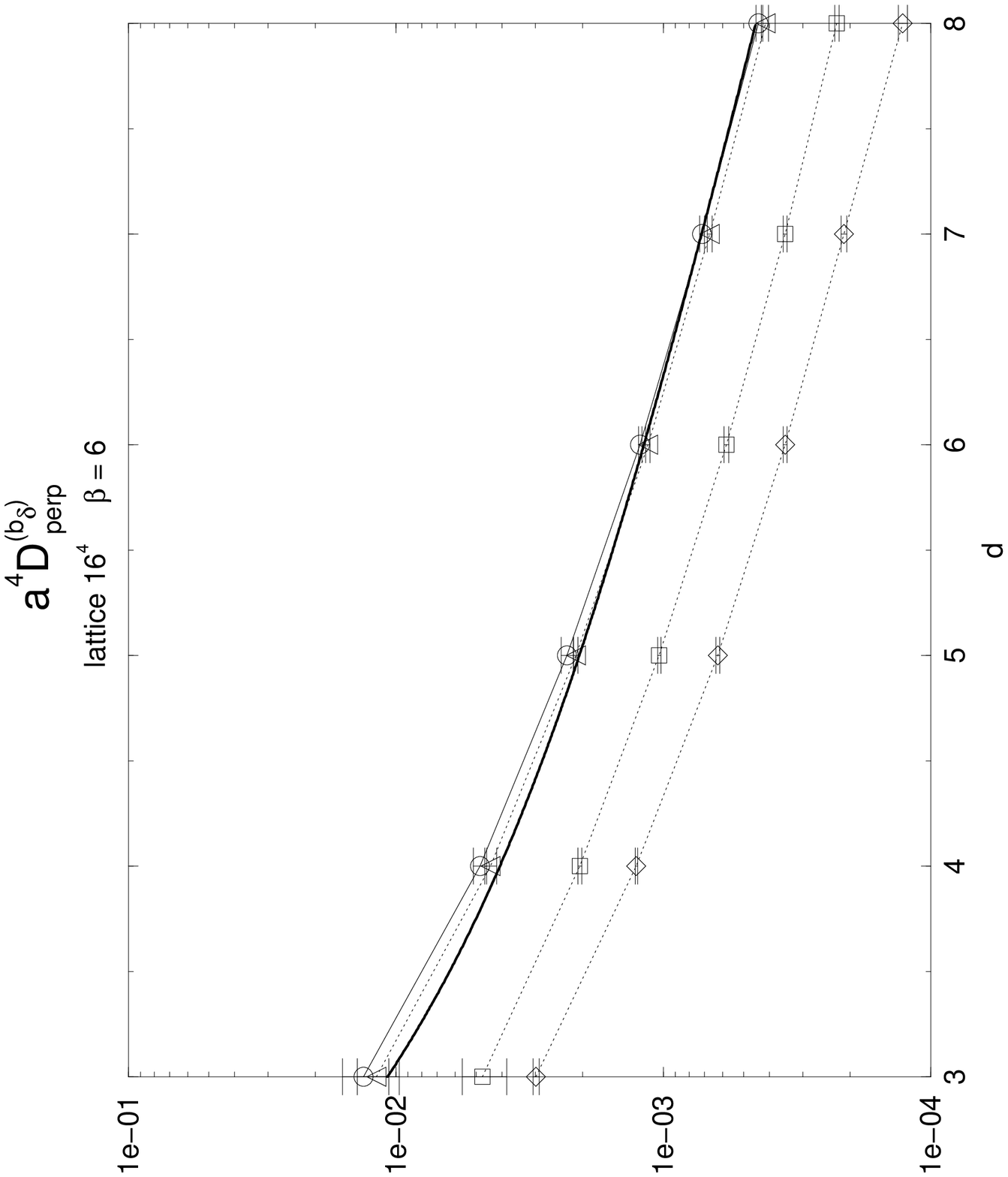}
\end{figure}

\vfill\eject

\centerline{\bf Figure 4}
\vskip 4truecm
\begin{figure}[htb]
\vskip 4.5truecm
\includegraphics{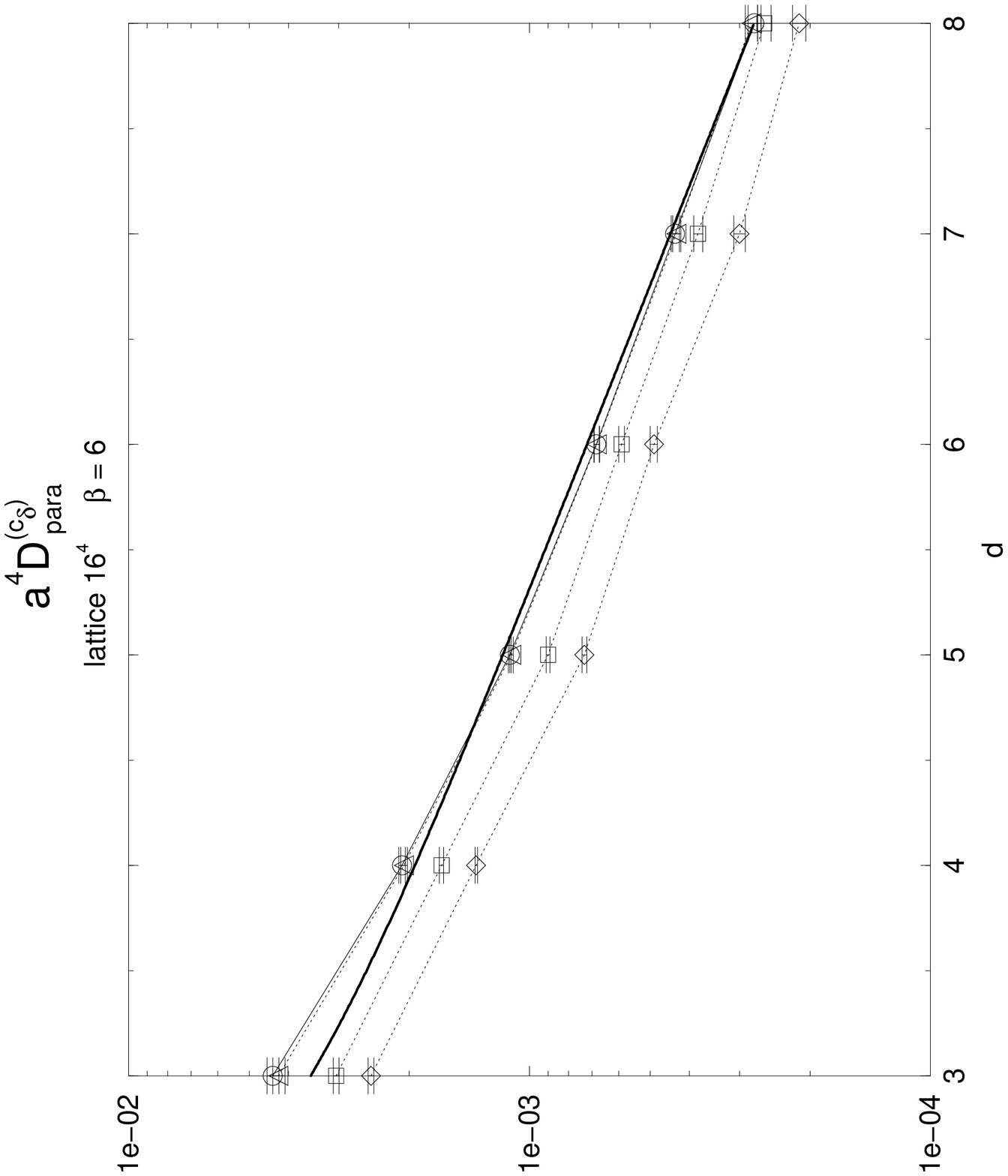}
\end{figure}

\vfill\eject

\centerline{\bf Figure 5}
\vskip 4truecm
\begin{figure}[htb]
\vskip 4.5truecm
\includegraphics{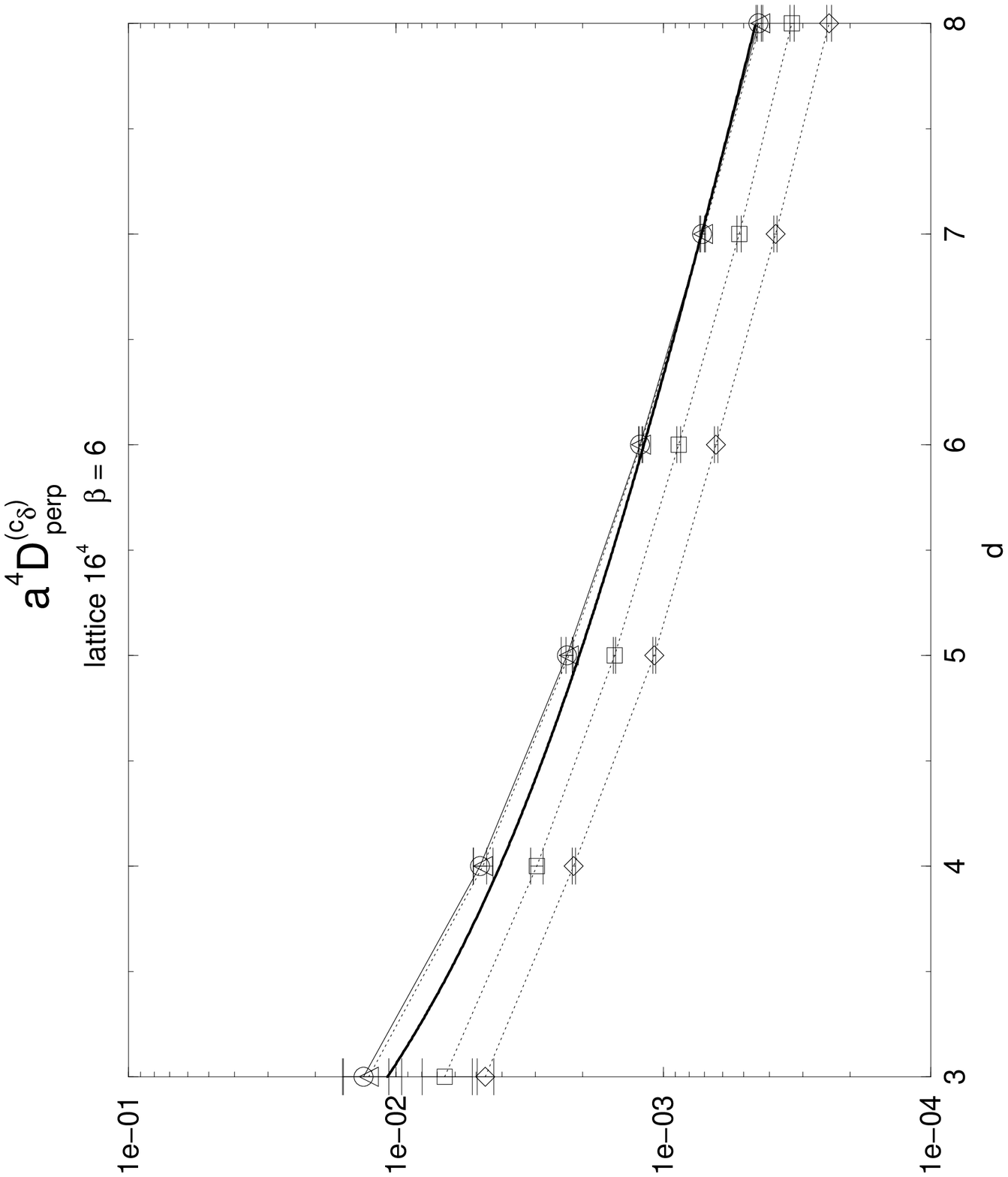}
\end{figure}

\vfill\eject

\end{document}